\documentclass[preprint,aps,showpacs,prb,amsmath,amssymb,floatfix]{revtex4-1}
\usepackage{graphicx}
\usepackage{dcolumn}
\usepackage{bm}
\usepackage{rotating}
\usepackage{float}
\usepackage{color}
\usepackage{subfigure}

\begin{document}

\date{\today} 
\title{
      First-principles study of structural, magnetic and electronic properties of small FeRh alloy clusters
      }
\author{Junais Habeeb Mokkath}\author{G. M. Pastor} 
 \affiliation{%
Institut f{\"u}r Theoretische Physik, 
Universit{\"a}t Kassel, 
Heinrich Plett Stra{\ss}e 40, 34132 Kassel, Germany
      }
\date{\today}

\begin{abstract}
The structural, electronic and magnetic properties of small 
${\rm Fe}_m {\rm Rh}_n$ clusters having $N = m+n \leq 8$ atoms 
are studied in the framework of a generalized-gradient approximation 
to density-functional theory.
For $N = m+n \leq 6$ a thorough sampling
of all cluster topologies has been performed, while for $N = 7$ and $8$  
only a few representative topologies are considered. In all cases the entire 
concentration range is systematically investigated.
All the clusters show ferromagnetic-like order in the optimized structures. As a result, the average magnetic moment per atom $\overline\mu_N$ increases 
monotonously, which is almost linear over a wide range of concentration with Fe content.
A remarkable enhancement of the local Fe moments beyond 3~$\mu_B$ is
observed as result of Rh doping. The composition 
dependence of the binding energy, average magnetic moment and electronic structure are discussed.

\end{abstract}
%

%

\maketitle

\section{Introduction}
\label{sec:introd}

Alloying elements with complementary qualities in order to tailor 
their physical behavior for specific technological purposes
has been a major route in material development since 
the antiquity. Cluster research is no exception to this trend. 
After decades of systematic studies of the size and structural
dependence of the most wide variety of properties of monoelement 
particles, the interest has actually been moving progressively 
over the past years towards investigations on finite-size binary 
alloys.\cite{faraday_vol}  The magnetism of transition-metal (TM) 
clusters opens numerous possibilities and challenges in this 
context.\cite{zitoun02,dennlerCoRh,IrenaPdCu,mukulCoMn,peter09,Antonis04,georg08,bansmann,antoniak06,yin,knick07,%
farle08,gruner08,four08,four09,dupuis08} 
For example, one would like to understand
how to modify the magnetic characteristics of clusters, in particular 
the saturation magnetization and the magnetic anisotropy energy (MAE),
as it has been done in solids. This would indeed
allow one to design new nanostructured materials from a
microscopic perspective. Nevertheless, it also true that 
controlling composition, system size, and magnetic  
behavior sets serious difficulties for both experiment and theory. 

Pure TM clusters such as Fe$_N$, Co$_N$ and Ni$_N$ show spin moments, 
orbital moments, and MAEs that are enhanced with respect to the 
corresponding periodic 
solids.\cite{pastor89,bucher91,billas93,prl_mae,apsel96,knick01,prb_gin} 
Still, the possibilities of optimizing the cluster magnetic behavior 
by simply tuning the system size have been rather disappointing, 
particularly concerning the MAE, which remains relatively small 
---despite being orders of magnitude larger than in solids\cite{prl_mae}--- 
due to the rather weak spin-orbit (SO) coupling in the $3d$ atoms. 
This is one of the motivations for alloying $3d$ TMs with $4d$ and 
$5d$ elements which, being heavier, are subject to stronger SO 
interactions. In this context it is useful to recall that large 
nanoparticles and three dimensional 
solids of these elements are non-magnetic. However, at very 
small sizes the $4d$ and $5d$ clusters often develop a finite spontaneous
low-temperature magnetization, due to the reduction of local 
coordination and the resulting $d$-band 
narrowing.\cite{galicia,onset,coxRh,reddy,pedroRh}
The first experimental observation of this important finite-size 
effect has been made by Cox {\em et al.}
by performing Stern-Gerlach-deflection measurements 
on Rh$_N$ clusters. In this work the average magnetic moments per 
atom $\overline\mu_N$ = $0.15$--$0.80 \mu_B$ have been experimentally determined
for $N \le 30$--$50$ atoms.\cite{coxRh} In view of these contrasting 
features one expects that $3d$-$4d$ and $3d$-$5d$
alloy clusters should show very interesting 
structural, electronic and magnetic behaviors. 

The purpose of this paper is to investigate the ground-state properties 
of the small FeRh clusters in the framework of Hohenberg-Kohn-Sham's 
density functional 
theory.\cite{hks} Besides the general interest of the problem 
from the perspective of $3d$-$4d$ nanomagnetism, these clusters
are particularly appealing because of the remarkable phase diagram 
of FeRh bulk alloys.\cite{phdiag} In the case of ${\rm Fe}_{50} {\rm Rh}_{50}$
the magnetic order at normal pressure and low temperatures is 
antiferromagnetic (AF). As the temperature increases this $\alpha''$ 
phase undergoes a first order transition to a ferromagnetic (FM) 
state, the $\alpha'$ phase, which is accompanied by a change in 
lattice parameter. The corresponding transition temperature 
$T_c^{\alpha'\alpha''}$ increases rapidly with increasing external 
pressure P, eventually displacing the FM $\alpha'$ phase completely 
for $P\ge 7$~GPa 
($T_c^{\alpha'\alpha''}\simeq290 K$ for ${\rm Fe}_{50} {\rm Rh}_{50}$ 
at normal pressure). Moreover, $T_c^{\alpha'\alpha''}$ decreases very 
rapidly with decreasing Rh content. 
At low pressures the FM $\alpha'$ phase undergoes a FM 
to paramagnetic (PM) transition at ($T_C \simeq 670 K$).\cite{phdiag} 
In addition, the properties of $\alpha$-FeRh bulk alloys have been the subject 
of first principles and model theoretical investigations.\cite{Entel_FeRh}
In particular these show that the relative stability of the
FM and AF solutions depends strongly on the interatomic distances.
Such remarkable condensed-matter effects enhance the
appeal of small FeRh particles as specific example of
$3d$-$4d$ nanoscale alloy. Investigations of their magnetic properties as a function of 
size, composition, and structure are therefore of fundamental
importance.

The remainder of the paper is organized as follows. 
In Sec.~\ref{sec:teo} the main details of the theoretical background
and computational procedure are presented. This includes in particular 
a description of the strategy used for exploring the cluster energy 
landscape as a function of geometrical conformation and chemical order. 
The results of our calculations for FeRh clusters having $N\le 8$ 
atoms are reported in Sec.~\ref{sec:stmagn} by analyzing
the concentration dependence of the cohesive energy, 
the local and average magnetic moments, and the spin-polarized electronic
structure. Finally, we conclude in Sec.~\ref{sec:concl} with a summary of 
the main trends and an outlook to future extensions.

\section{Computational aspects}
\label{sec:teo}

The calculations reported in this work are based on density functional theory,\cite{hks}
as implemented in the Vienna ab initio simulation package (VASP).\cite{vasp} 
The exchange and correlation energy is described by using both the spin-polarized 
local density approximation (LDA) and 
Perdew and Wang's generalized-gradient approximation (GGA).\cite{pw91} 
The VASP solves the spin-polarized Kohn-Sham equations in an augmented 
plane-wave basis set, taking into account the core electrons within the 
projector augmented wave (PAW) method.\cite{paw} 
This is an efficient frozen-core all-electron approach which allows to 
incorporate the proper nodes of the Kohn-Sham orbitals in the core region
and the resulting effects on the electronic structure, total energy and 
interatomic forces. The $4s$ and $3d$ orbitals of Fe, and the $5s$ and 
$4d$ orbitals of Rh are treated as valence states. The wave functions 
are expanded in a plane wave basis set with the kinetic energy cut-off $E_{max}=268$~eV. 
In order to improve the convergence of the solution of the selfconsistent 
KS equations the discrete energy levels are broadened by using a Gaussian smearing
$\sigma = 0.02$~eV. The validity of the present choice of computational parameters 
has been verified.\cite{foot-acc} The PAW sphere radii for Fe and Rh are $1.302$~{\AA} and $1.402$~{\AA}, 
respectively. A simple cubic supercell is considered with 
the usual periodic boundary conditions. The linear size of the cell is $a$ = $10$--$22$~{\AA}, so that 
any pair of images of the clusters are well separated and 
the interaction between them is negligible.
Since we are interested in finite systems, the reciprocal space summations are restricted 
to the $\Gamma$ point.

Although the potential advantages of alloying magnetic $3d$ elements 
with highly-polarizable $4d$ or $5d$ elements 
are easy to understand, the problem 
involves a number of serious practical challenges.
Different growth or synthesis conditions can lead to different 
chemical orders, which can be governed not just by energetic reasons
but by kinetic processes as well. For instance, one may have to deal with
segregated clusters having either a $4d$ core and a $3d$ outer shell or 
vice versa. Post-synthesis manipulations can induce different 
degrees of intermixing, including for example surface diffusion 
or disordered alloys. 
Moreover, the inter atomic distances are also expected to 
depend strongly on size and composition. Typical TM-cluster bond-lengths 
are in fact $10$--$20$\% smaller than in the corresponding bulk crystals. 
Taking into account that itinerant $\rm 3d$-electron magnetism
is most sensitive to the local and chemical environments of the 
atoms,\cite{onset,faradayCoRh,phm,garyclus} 
it is clear that controlling the distribution of the elements 
within the cluster   
is crucial for understanding magnetic nanoalloys.

Systematic theoretical studies of binary-metal clusters are hindered by the
diversity of geometrical conformations, ordered and disorder arrangements, 
as well as segregation tendencies that have to be taken into account. This 
poses a serious challenge to both first-principles and model 
approaches. In order to determine the interplay between cluster structure, 
chemical order and magnetism in FeRh clusters we have performed a 
comprehensive set of electronic calculations for clusters having 
$N \le 8$ atoms. In the present paper we focus on the most stable 
cluster structure and magnetic configuration,
which are determined by exploring the ground-state energy landscape.\cite{tobe}
This is a formidable task, since one needs to consider a large, most possibly complete and unbiased set of 
initial structures. Such a thorough geometry optimization must include not only the representative cluster
geometries or topologies, but also all relevant chemical orders. This requires taking into account
all distributions of the Fe and Rh atoms for any given size and composition.
These two aspects of the problem of determining the structure of nanoalloys are discussed in more detail in the following. 

The different cluster topologies are sampled
by generating all possible graphs for $\rm N\le 6$ 
atoms as described in Ref.~\onlinecite{phm} (see also Ref.~\onlinecite{Wang}).
For each graph or adjacency matrix it is important to verify 
that it can be represented by a true structure in $\rm D\le 3$ dimensions. 
A graph is acceptable as a cluster structure, only if a set of atomic 
coordinates $\vec R_i$ with $i = 1, \dots, N$ exists, such that the 
interatomic distances $R_{ij}$ satisfy the conditions
$R_{ij}=R_0$ if the sites $i$ and $j$ are connected in the graph 
(i.e., if the adjacency matrix element $A_{ij} =1$) 
and $R_{ij}> R_0$ otherwise (i.e., if $A_{ij} =0$). 
Here $R_0$ refers to the nearest neighbor (NN) distance, which 
at this stage can be regarded as the unit of length, assuming for 
simplicity that it is the same for all clusters.
Notice that for $N\le 4$ all graphs are possible cluster 
structures. For example, for $N = 4$, the different structures are the tetrahedron, rhombus,
square, star, triangular racket and linear chain.\cite{phm} However, for $N\ge 5$ there are graphs, i.e., topologies, which cannot
be realized in practice. For instance, it is not possible to have five 
atoms being NNs from each other in a three dimensional space. Consequently, for $N\ge 5$ there are less real structures than mathematical graphs. 
The total number of graphs (structures) is 21 (20), 112 (104), and 853 (647)
for $N = 5, 6$, and $7$, respectively.\cite{phm} 

For clusters having $N\le 6$ atoms all these topologies have
indeed been taken as starting points of our structural relaxations.
Out of this large number of different initial configurations 
the unconstrained relaxations using VASP lead to only a few geometries, 
which can be regarded as stable or metastable isomers. For larger clusters
($N = 7$ and $8$) we do not aim at performing a full global optimization. Our purpose here is to explore
the interplay between magnetism and chemical order as a function of composition
for a few topologies that are representative of open and close-packed structures. Taking
into account our results for smaller sizes, and the available information
on the structure of pure Fe$_N$ and Rh$_N$ clusters, we have restricted the 
set of starting topologies for the unconstrained relaxation of FeRh
heptamers and octamers to the following: bicapped trigonal bipyramid, capped octahedra, and pentagonal bipyramid for $N = 7$, and
tricapped trigonal bipyramid, bicapped octahedra, capped pentagonal bipyramid 
and cube for $N = 8$. Although, the choice of topologies for $N = 7$ and $8$ is quite restricted, it includes
compact as well as more open structures. Therefore, it is expected to shed light on the dependence 
of the magnetic properties on the chemical order and composition.

The dependence on concentration is investigated systematically for each topology  
of Fe$_m$Rh$_n$ by varying $m$ and for each size $N = m+n \leq 8$, including the pure Fe$_N$ and Rh$_N$ limits. Moreover, we 
take into account all possible non-equivalent distributions of the $m$ Fe 
and $n$ Rh atoms within the cluster. In this way, any {\em a priori}
assumption on the chemical order is avoided. Obviously, such an
exhaustive combinatorial search increasingly complicates the computational 
task as we increase the cluster size, and as we move away from pure clusters 
towards alloys with equal concentrations. Finally, in order to
perform the actual density-functional calculations we set for 
simplicity all NN distances in the starting cluster geometry equal to the Fe bulk  
value\cite{foot_R0} $R_0 = 2.48$~{\AA}. Subsequently, a fully unconstrained geometry optimization is performed from 
first principles by using the VASP.\cite{vasp} 
The atomic positions are fully relaxed by means of conjugate gradient 
or quasi-Newtonian methods, without imposing any symmetry constraints, 
until all the force components are smaller than the threshold
$5$~meV/{\AA}. The convergence criteria are set 
to $10^{-5}$~eV/{\AA} for the energy gradient, and 
$5\times10^{-4}$~{\AA} for the atomic displacements.\cite{vasp_manual} 
The same procedure applies to all considered clusters 
regardless of composition, chemical order, or total magnetic moment.
Notice that the diversity of geometrical structures and atomic arrangements 
often yields many local minima on the ground-state energy surface, which complicates 
significantly the location of the lowest-energy configuration.

Lattice structure and magnetic behavior are intimately related in TMs,
particularly in weak ferromagnets such as Fe and its alloys.\cite{foot_struct}
On the one side, the optimum structure and chemical order depend on the
actual magnetic state of the cluster as given by the average magnetic 
moment per atom $\overline\mu_N$ and the magnetic order. On the other side, 
the magnetic behavior is known to be different for different structures
and concentrations. Therefore, in order to rigorously determine the 
ground-state magnetic properties of FeRh clusters, we have varied systematically the value of the total 
spin polarization of the cluster $S_z$ by performing fixed spin-moment (FSM) calculations in the whole physically 
relevant range. Let us recall that $S_z = (\nu_\uparrow - \nu_\downarrow)/2$
where $\nu_\uparrow$($\nu_\downarrow$) represents the number of electrons in the majority (minority)
states. In practice we start from the non-magnetic 
state ($S_z^{min} = 0$) and increase $S_z$ until the local spin moments
are fully saturated, i.e., until the Fe moments in the PAW sphere
reach $\mu_{\rm Fe}\simeq 4 \mu_B$ and the Rh moments 
$\mu_{\rm Rh}\simeq 2.5 \mu_B$ (typically, $S^{max}_{z} \gtrsim 3N/2$).
The above described global geometry optimizations  are performed 
independently for each value of $S_z$. These FSM study
provides a wealth of information on the isomerization energies, the spin-excitation energies, and their 
interplay. These are particularly interesting for a 
subtle magnetic alloy such as FeRh, and would therefore deserve to be analyzed in some more detail. 
In the present paper we shall focus
on the ground-state properties by determining for each considered 
Fe$_m$Rh$_n$ the most stable structural and magnetic
configuration corresponding to energy minimum as a function of $S_z$
and of the atomic positions.\cite{tobe}

Once the optimization with respect to structural and magnetic degrees of 
freedom is achieved, we derive the binding energy per atom 
$E_B = [m E({\rm Fe}) + n E({\rm Rh}) - E({\rm Fe}_m {\rm Rh}_n) ]/ N$
in the usual way by referring the total energy $E$ to the corresponding 
energy of $m$ Fe and $n$ Rh isolated atoms. Moreover, for each stationary 
point of the total energy surface (i.e., for each relaxed structure having 
a nearly vanishing $\|{\vec{\nabla}E}\|$)
we determine the vibrational frequencies from the diagonalization of 
the dynamical matrix. The latter is calculated from finite differences 
of the analytic gradients of the total energy. In this way we can 
rule out saddle points to which the local optimization procedure
happens to converge on some occasions. Only configurations which 
correspond to true minima are discussed in the following. 
Finally, a number of electronic and magnetic properties 
---for example, the binding energy, the local 
magnetic moments $\mu_{i}$ integrated within the Wigner-Seitz (WS) or Bader atomic 
cells of atom $i$,\cite{bader,thesis_jl} and the spin polarized 
density of electronic states (DOS) 
$\rho_{\sigma}(\varepsilon)$--- are derived from the self-consistent
spin-polarized density and Kohn-Sham spectrum.

\section{Results and discussion}
\label{sec:stmagn}
In the following we present and discuss results for the binding energy, average and local spin 
moments, and electronic densities of states for $N = m + n \le 8$.

\subsection{Binding energy and magnetic moments} 
\label{sec:EB}

\begin{figure}
\begin{center}
  \includegraphics[width=8cm,height=5.2cm,angle=0,clip=true]{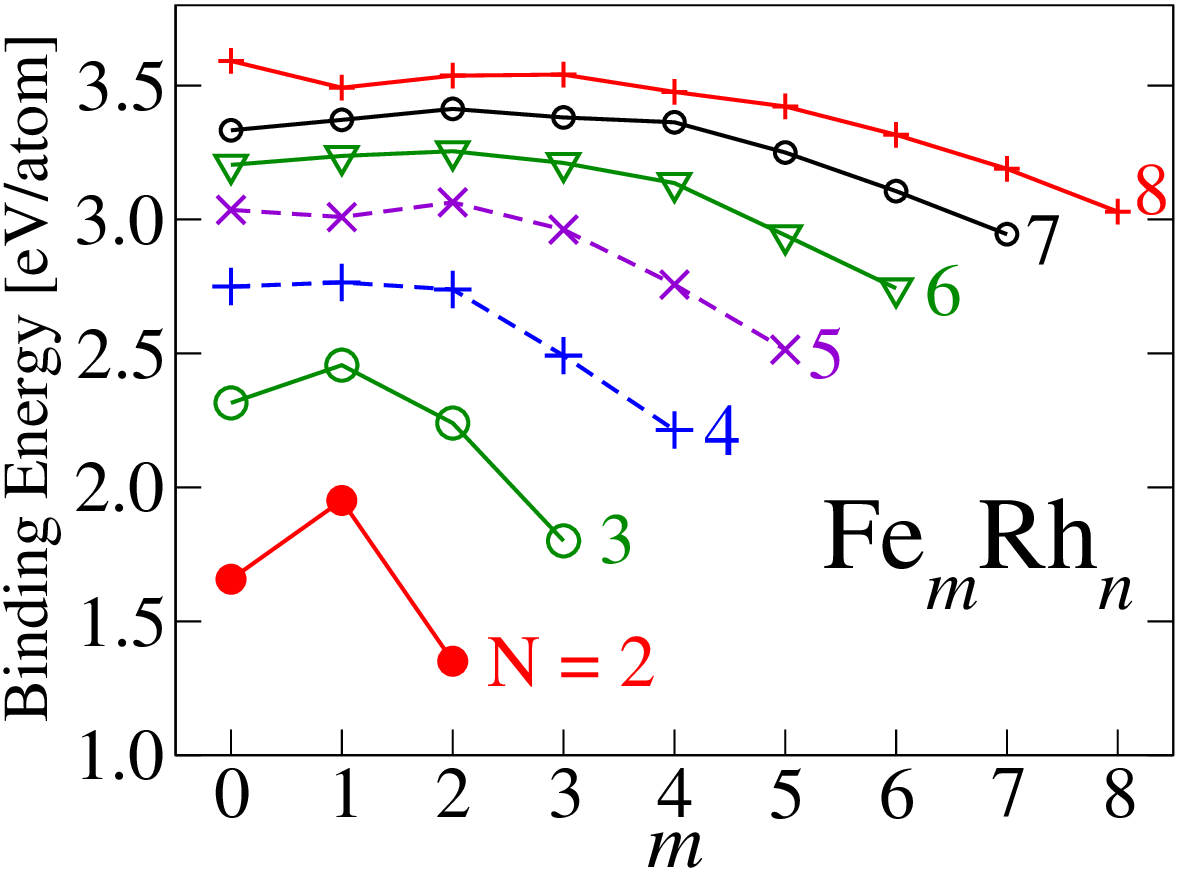}
\end{center}
\caption{(Color online) 
Binding energy per atom $E_B$ of Fe$_m$Rh$_n$ 
clusters as a function of the number of Fe atoms. The lines connecting 
the points for each $N = m + n$ are a guide 
to the eye.
        }
\label{fig:EB}
\end{figure}

In Fig.~\ref{fig:EB} the binding energy per atom $E_B$ is given as
a function of the number of Fe atoms $m$. Besides the expected 
monotonic increase of $E_B$ with increasing $N$, an interesting 
concentration dependence is observed. For very small sizes
($N\le 4$) $E_B$ is maximal for $m=1$ or $2$, despite the fact
that $E_B$ is always larger for pure Rh than pure Fe clusters.
This indicates that in these cases the bonding resulting
from FeRh pairs is stronger than RhRh bonds. Only for $m \ge  N - 1$,
when the number of weaker FeFe bonds dominates, one observes that 
$E_B$ decreases with increasing $m$. For larger sizes ($N\ge 5$)
the strength of RhRh and FeRh bonds becomes very similar, so that 
the maximum in $E_B$ is replaced by a range of
Fe concentrations $x = m /N \lesssim 0.5$ where $E_B$ depends
very weakly on $m$.

\begin{figure}
\includegraphics[width=8cm,height=5.2cm,angle=0,clip=true]
                {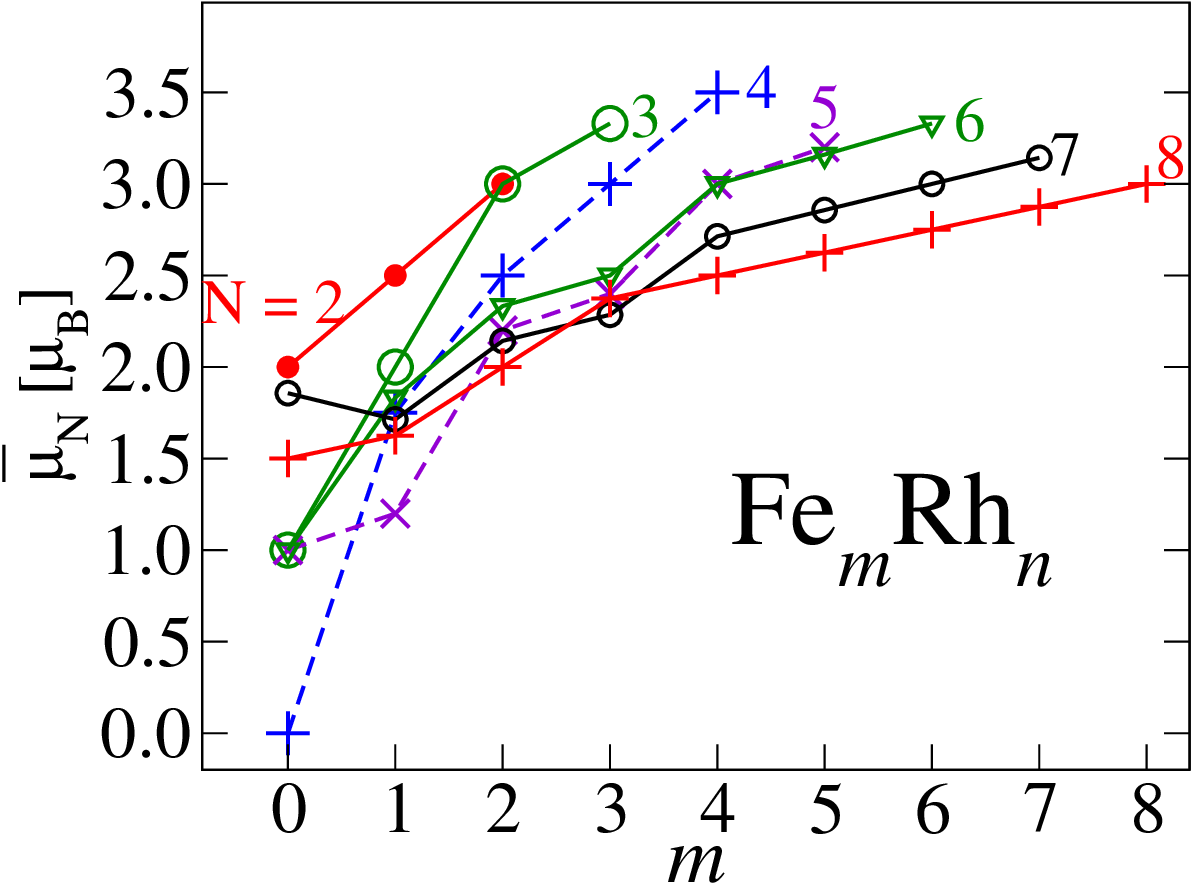}
\caption{(Color online) 
Total magnetic moment per atom $\overline\mu_{N}$ of Fe$_m$Rh$_n$ 
clusters as a function of number of Fe atoms. The symbols corresponding to each size are the same as in Fig.~\ref{fig:EB}.
The lines connecting the points for each $N = m + n$ are a guide to the eye. 
        }
\label{fig:moma}
\end{figure}

In Fig.~\ref{fig:moma} the average magnetic moments $\overline\mu_N$
of Fe$_m$Rh$_n$ are shown as a function of $m$ for $N\le 8$.
First of all, one observes that $\overline\mu_N$ increases 
monotonously, with the number of Fe atoms. 
This is an expected consequence of the larger Fe local moments and
the underlying FM-like magnetic order. 
The average slope of the curves tends to increase with decreasing $N$, since 
the change in concentration per Fe substitution is more important the 
smaller the size is. The typical increase in $\overline\mu_N$
per Fe substitution is about ($1/N$)$\mu_B$ per Fe substitution.
Notice, moreover, the enhancement of the magnetic moments of the pure 
clusters in particular for Fe$_N$ ($m=N$), which go well beyond 
$3 \mu_B$, the value corresponding to a saturated $d$-band in the $d^7s^1$ 
configuration. In contrast, the moments of pure Rh$_N$ are far from saturated except for $N = 2$ and $7$ 
(see Fig.~\ref{fig:moma} for $m=0$). 
In this context it is important to recall that a thorough global optimization, for example, by considering a large number
of initial topologies, could affect the quantitative values of the magnetic moments for $N = 7$ and $8$.

\begin{figure}
\includegraphics[width=8cm,height=5.2cm,angle=0,clip=true]
                {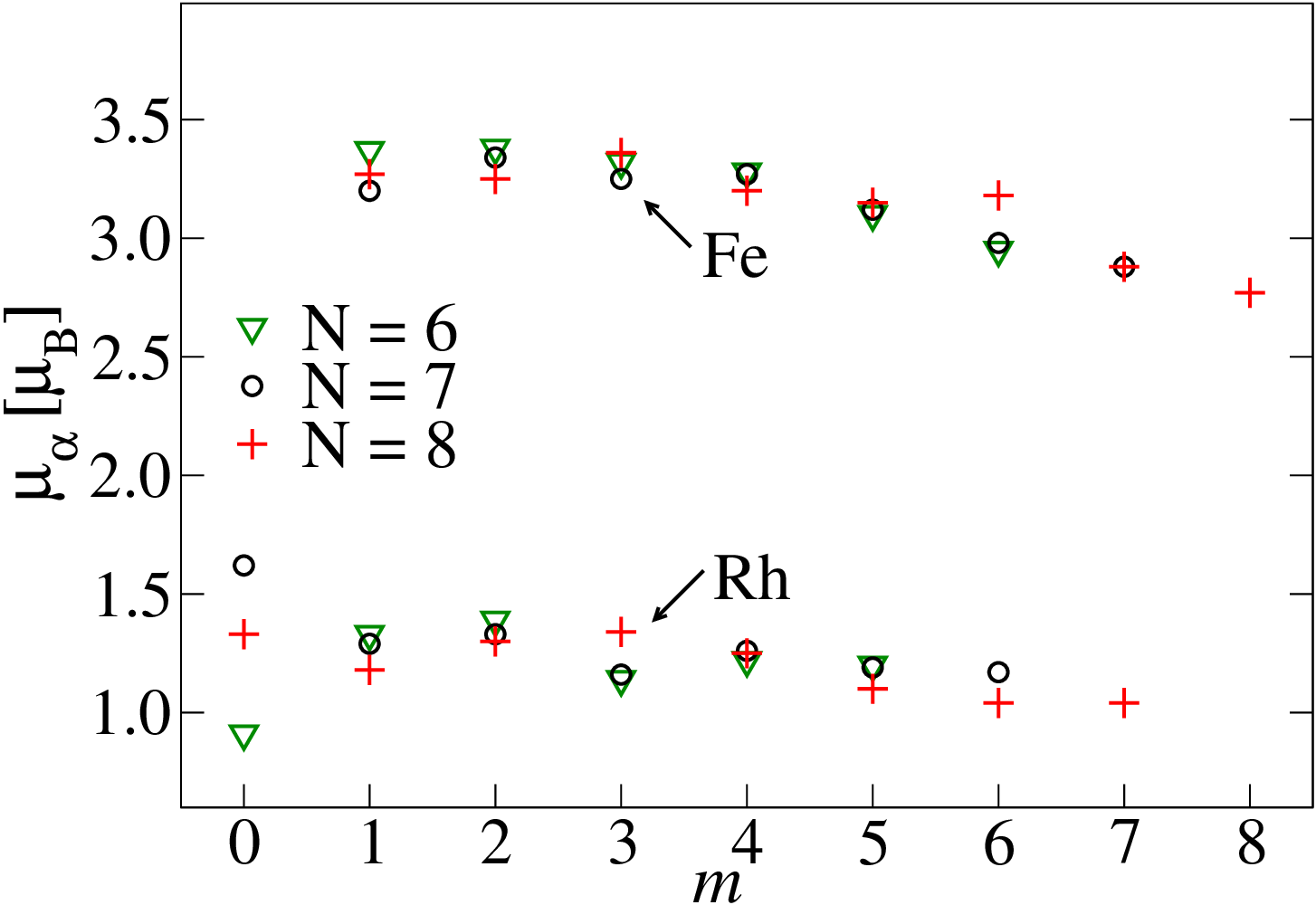}
\caption{(Color online) 
Local magnetic moment $\mu_{\alpha}$ at the Fe and Rh atoms as a function 
of the number Fe atoms $m$.}
\label{fig:moml}
\end{figure}

The local magnetic moments in the PAW sphere of the Fe and Rh atoms 
provide further insight on the interplay between $3d$ and $4d$ 
magnetism in Fe$_m$Rh$_n$. In Fig.~\ref{fig:moml} $\mu_{\rm Fe}$ and $\mu_{\rm Rh}$ are shown as a function 
of $m$ for $N = 6$--$8$. The Fe moments are 
essentially given by the saturated $d$-orbital contribution.
For pure Fe clusters the actual values of $\mu_{\rm Fe}$ within the PAW sphere 
are somewhat lower than $3\mu_B$ due to a partial spill-off of
the spin-polarized density. Notice that the Fe moments increase as
we replace Fe by Rh atoms showing some weak oscillations as a function of $m$.
The increase is rather weak for a single
Rh impurity in Fe$_{N-1}$Rh but becomes stronger reaching a more or less
constant value as soon as the cluster contains 2 or more 
Rh ($m \le N-2$, see Fig.~\ref{fig:moml}). This effect can be traced 
back to a $d$ electron charge transfer from Fe to Rh which, together with
the extremely low coordination number, yields a full polarization of the 
larger number of Fe $d$ holes. On the other side the Rh moments 
are not saturated and therefore are more sensitive to size, structure
and composition. The values of $\mu_{\rm Rh}$ are in the range of 
$1$--$1.5\mu_B$ showing some oscillations as a function of $m$. 
No systematic enhancement of $\mu_{\rm Rh}$ with increasing Fe content
is observed. This behavior could be related to charge transfers effects
leading to changes in the number of Rh $d$ electrons as a function of $m$.

Finally, it is interesting to analyze the role played by
magnetism in defining the cluster structure by comparing
magnetic and non-magnetic calculations. For the smallest
FeRh clusters ($N=3$ and $4$) the magnetic energy $\Delta E_m = E(S_z \!=\!0) - E(S_Z)$ 
gained upon magnetization is higher in the first excited isomer than in
the most stable structure. This implies that the contribution of magnetism
to the structural stability is not crucial, since the non-magnetic calculations
yield the same ordering, at least concerning the
two best structures. This suggests that for the smallest sizes
the kinetic or bonding energy dominates the structural stability,
which also explains that the two most stable isomers
have different topologies.
The situation changes for large clusters. For $N\ge 5$ one finds a number of
FeRh clusters for which the optimal structure is actually stabilized by magnetism.
For example, in Fe$_4$Rh, Fe$_3$Rh$_2$, and FeRh$_4$ the energy ordering
of the two most stable isomers would be reversed if magnetism were neglected.
It should be noted that in these cases the structures differ only in the 
chemical order, not in the topology which is a TBP. In the FeRh hexamers 
the energy differences between the low-lying isomers are more important and only
in one case, Fe$_4$Rh$_2$, magnetism appears to be crucial for stabilizing the
actual optimal structure. A similar strong interplay between structure, 
chemical order and magnetism is expected for larger FeRh clusters.

\subsection{Electronic structure} 
\label{sec:elst}

\begin{figure*}
\includegraphics[width=15cm,height=13.6cm,angle=0,clip=true]
                {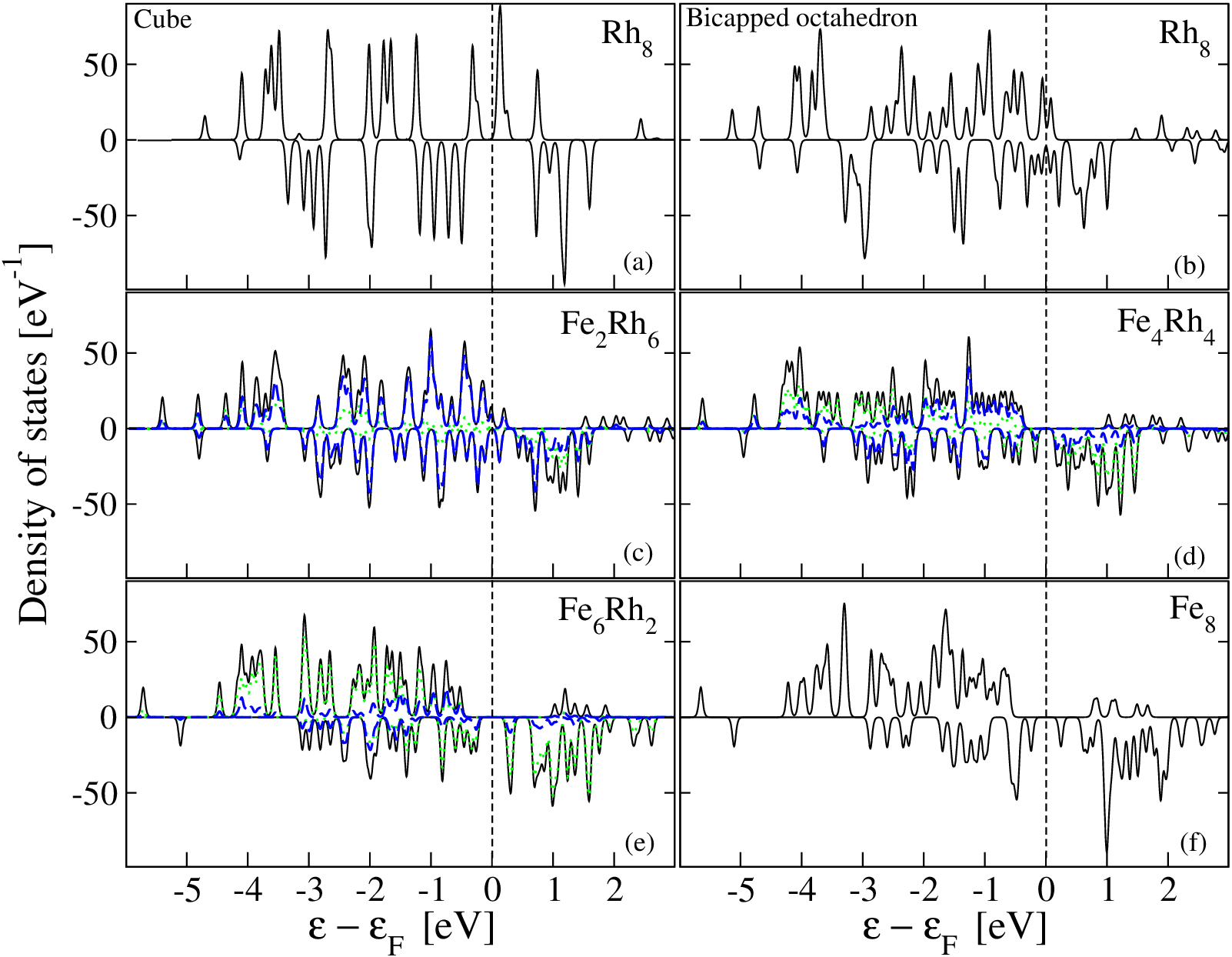}
\caption{(Color online) 
Electronic density of states (DOS) of FeRh octamers.
Results are given for the total (solid), the Fe-projected (dotted), 
and the Rh-projected (dashed) $d$-electron DOS. Positive (negative) 
values correspond to majority (minority) spin. 
A Lorentzian width $\lambda = 0.02$~eV has been used to
broaden the discrete energy levels. 
        }
\label{fig:dos}
\end{figure*}

It is very interesting to analyze, at least for some representative examples, how the 
electronic structure depends on the composition of magnetic 
nanoalloys. To this aim we report in Fig.~\ref{fig:dos}
the spin-polarized $d$-electron density of states (DOS) of 
representative FeRh octamers having the 
most stable relaxed configuration among the considered topologies (see Sec.~\ref{sec:teo}).
Results for pure Fe$_8$ and Rh$_8$ are also shown for the sake of comparison.
In all the clusters, the dominant peaks in the relevant energy range
near $\varepsilon_F$ correspond either to the Fe-$3d$ or to
the Rh-$4d$ states. The valence spectrum is 
largely dominated by these $d$-electron contributions. In fact the
total DOS and the $d$-projected DOS are difficult to tell apart. 

First of all, let us consider the DOS of the pure clusters. Our results for Rh$_8$ with a cube structure are similar to those 
of previous studies.\cite{vijay05} They show the dominant $d$-electron contribution near $\varepsilon_F$, with the characteristic  
ferromagnetic exchange splitting between the minority and majority spin states.
In Fig.~\ref{fig:dos} we also included the DOS for Rh$_8$ with a BCO structure, since it allows us
to illustrate the differences in the electronic structure of compact and 
open geometries. Moreover, the DOS of pure Rh$_8$ with BCO structure is very useful in order to demonstrate
the dependence of DOS on Fe content, since the structures of 
Fe$_m$Rh$_{8-m}$ with $m \geq 1$ are similar to the BCO. 
Both Fe$_8$ and Rh$_8$ show relatively narrow $d$-bands
which dominate the single-particle energy spectrum in the range 
$-5 {\rm eV} \le \varepsilon - \varepsilon_F \le 3 {\rm eV}$. 
The spin polarization of the DOS clearly reflects the ferromagnetic order
in the cluster. Putting aside the exchange splitting, the peak 
structure in the up and down DOS $\rho_\sigma(\varepsilon)$
are comparable. There are even  
qualitative similarities between the two elements. However, looking in 
more detail, one observes that the effective $d$-band width in
Fe$_8$ (about 4~eV) is somewhat smaller than in Rh$_8$ (about 5~eV).
Moreover, in Rh$_8$ the DOS at $\varepsilon_F$ is non-vanishing for both spin 
directions and the finite-size gaps are very small 
(see Fig.~\ref{fig:dos}). In contrast, the majority $d$-DOS 
is fully occupied in Fe$_8$, with the highest majority state
lying about 0.5~eV below $\varepsilon_F$. In addition there
is an appreciable gap (about 0.1~eV) in the corresponding 
minority spectrum. These qualitative differences are of course consistent
with the fact that Fe$_8$ is a strong ferromagnet with saturated moments, 
while Rh$_8$ should be regarded as a weak unsaturated ferromagnet.

The trends as a function of concentration reflect the crossover between
the previous contrasting behaviors.
For low Fe concentration (e.g., Fe$_2$Rh$_6$)
we still find states with both spin directions close to 
$\varepsilon_F$. The magnetic moments are not saturated, 
although the Fermi energy tends to approach the 
top of the majority band. Moreover, the majority-spin states 
close to $\varepsilon_F$ have dominantly Rh character. Small Fe doping
does not reduce the $d$-band width significantly. Notice the rather
important change in the shape of the DOS in Fe$_2$Rh$_6$ as compared
to the DOS in Rh$_8$. This is a consequence of the change in topology
from cubic to bicapped octahedron (BCO).  
 
For equal concentrations (Fe$_4$Rh$_4$) the first signs of 
$d$-band narrowing and enhanced exchange splitting 
start to become apparent. The spin-up states (majority band) 
which in Fe$_2$Rh$_6$ contribute to the DOS at $\varepsilon_F$ 
now move to lower energies (0.3 eV below $\varepsilon_F$) 
so that the majority band 
is saturated. Only spin-down (minority) states are found around $\varepsilon_F$,
although there is a significant gap in $\rho_\downarrow(\varepsilon)$  
(see Fig.~\ref{fig:dos}). In the majority band Rh dominates over Fe 
at the higher energies (closer to $\varepsilon_F$), while Fe dominates 
in the bottom of the band. In the minority band the participation of Rh (Fe)
is stronger (weaker) below $\varepsilon_F$ and weaker (stronger) 
above $\varepsilon_F$. This is consistent with the fact that the Rh 
local moments are smaller than the Fe moments. 

Finally, in the Fe rich limit (e.g., Fe$_6$Rh$_2$), 
the majority-band width becomes as narrow as in Fe$_8$,
while the minority band is still comparable to Rh$_8$.
The exchange splitting is large, the majority band  
saturated and only minority states are found close
to $\varepsilon_F$. As in Fe$_8$, $\rho_\downarrow(\varepsilon)$  
shows a clear gap at $\varepsilon_F$ (see Fig.~\ref{fig:dos}).
However, the Rh contribution to the minority states below
$\varepsilon_F$ remains above average despite the relative
small Rh content. The Fe contribution largely
dominates the unoccupied minority-spin DOS, in agreement with
the larger local Fe moments. 

\section{Summary and outlook}
\label{sec:concl}

The structural, electronic and magnetic properties of small  
Fe$_m$Rh$_n$ clusters having $N = m+n \le 8$ atoms have been investigated 
systematically in the framework of a generalized gradient approximation 
to density-functional theory.
For very small sizes ($N\le 4$ atoms) 
the binding energy $E_B$ shows a non-monotonous dependence on concentration,
which implies that the FeRh bonds are stronger than the homogeneous ones. 
However, for larger sizes the FeRh and RhRh bond strengths 
become comparable, so that $E_B$ depends weakly on concentration for
high Rh content.

The magnetic order of the clusters having the most stable structures is found to be FM-like. 
Moreover, the average magnetic moment per atom $\overline\mu_N$ increases 
monotonously, which is almost linear over a wide range of concentration with Fe content.
Consequently, the energy gain $\Delta E_m$ associated to magnetism also
increases with the number of Fe atoms. The largest part of the 
spin polarization (about 90\%) can be traced back to the 
local $d$ magnetic moments within the PAW sphere of the atoms.
The $s$ and $p$ spin polarizations are almost negligible in general.
A remarkable enhancement of the local Fe moments is
observed as result of Rh doping. This is a 
consequence of the increase in the number of Fe $d$ holes, 
due to charge transfer from Fe to Rh, combined with 
the extremely reduced local coordination.
The Rh local moments are important already in the
pure clusters ($N\le 8$). Therefore, they are not significantly enhanced
by Fe doping. However, the overall stability of magnetism,
as measured by the total energy gained by the onset of spin 
polarization, is found to increase with increasing Fe content.

FeRh clusters are expected to develop a variety of further interesting 
behaviors, which still remain to be explored. For instance, 
larger cluster should show a more complex dependence of the magnetic 
order as a function of concentration. In particular for large Rh content
one should observe a transition from FM-like to 
AF-like order with increasing cluster size, in agreement
with the AF phase found in solids for more than 50\% Rh concentration. 
Moreover, the metamagnetic transition observed in bulk FeRh alloys 
also puts forward
the possibility of similar interesting phenomena in nanoalloys as 
a function of temperature. Finally, the contributions of orbital 
magnetism and magnetic anisotropy deserve to be explored in 
detail as a function of composition and chemical order, even for 
the smallest sizes, particularly because of their implications 
for potential applications.\cite{apl}

\begin{acknowledgments}

It is pleasure to thank Dr.~J.~L.~Ricardo-Ch\'avez and Dr. L. D\'iaz-S\'anchez for helpful discussions
and useful comments. Computer resources provided by ITS (Kassel) and CSC (Frankfurt) are gratefully acknowleged.

\end{acknowledgments}

\begin{thebibliography}{99}
\bibitem{foot-acc}
A number of tests have been performed in order to assess the 
numerical accuracy of the calculations. Increasing the cut-off energy E$_{max}$ = 268 eV and supercell size $a$ = 12~{\AA} to E$_{max}$ = 500 eV and $a$ = 22~{\AA}
in Rh$_4$ increases the computation time
by a factor $4$--$7$. This yields a total energy differences of 1.75 meV
and 0.25 meV, respectively. In the above calculations the changes in average 
bond length (bond angle) amounts to 10$^{-3}$ {\AA} (10$^{-4}$ degrees). These difference are not significant 
for our physical conclusions. In fact, typical isomerization energies in these clusters 
are an order of magnitude larger, of the order of $10$--$30$ meV. We also found that the total energy 
is nearly independent of the choice of the smearing parameter $\sigma$, provided it is not too large ($\sigma$ $\leq$ 0.05~eV). Values from $\sigma$ = 0.01 to 0.1 eV have been checked. Therefore, we judge
that our set of standard parameters (E$_{max}$ = 268 eV,
supercell size $a$ from 10 to 22 {\AA}, and $\sigma$ = 0.02 eV)
offers a sufficiently good accuracy at a reasonable computational costs.
\bibitem{tobe}
Results for lower lying isomers and excited magnetic configurations
will be reported elsewhere.
\bibitem{foot_R0}
The precise choice of the NN distances is not very important, since this
concerns merely the configuration for starting the unconstrained 
structural relaxation.
\end{thebibliography}
\end{document}